\documentclass[letter]{jpsj2} 
\title{Magnetic-field-induced enhancement of the vortex pinning in the overdoped regime of La$_{2-x}$Sr$_x$CuO$_4$ : Relation to the microscopic phase separation}

\author{\textsc{Yoichi Tanabe}\thanks{E-mail address: youichi@teion.apph.tohoku.ac.jp}, \textsc{Tadashi Adachi}, \textsc{Keisuke Omori}, \textsc{Hidetaka Sato}, \textsc{Takashi Noji} and \textsc{Yoji Koike}}

\inst{Department of Applied Physics, Graduate School of Engineering, Tohoku University, \\6-6-05 Aoba, Aramaki, Aoba-ku, Sendai 980-8579}

\abst{In order to investigate the inhomogeneity of the superconducting (SC) state in the overdoped high-$T_{\rm c}$ cuprates, we have measured the magnetic susceptibility, $\chi$, of La$_{2-x}$Sr$_x$CuO$_4$ (LSCO) single crystals in the overdoped regime in magnetic fields parallel to the c-axis up to 7 T on warming after zero-field cooling.
It has been found for $x$ = 0.198 and 0.219 that the temperature dependence of $\chi$ in 1 T shows a plateau, that is, $\chi$ is almost independent of temperature in a moderate temperature-range in the SC state.
Moreover, a so-called second peak in the magnetization curve has markedly appeared in these crystals.
These results indicate an anomalous enhancement of the vortex pinning and strongly suggest the occurrence of a $microscopic$ phase separation into SC and normal-state regions in the overdoped high-$T_{\rm c}$ cuprates.}

\kword{phase separation, La$_{2-x}$Sr$_x$CuO$_4$, magnetic susceptibility, second-peak effect}

\begin{document}
\maketitle

\section{Introduction}\label{intro}
Recently, the electronic inhomogeneity in the overdoped high-$T_{\rm c}$ cuprates has attracted interest in relation to the mechanism of the high-$T_{\rm c}$ superconductivity. 
Early studies of the specific heat of La$_{2-x}$Sr$_x$CuO$_4$ (LSCO) and Tl$_2$Ba$_2$CuO$_{6+\delta}$ (TBCO) have revealed that the electronic specific-heat coefficient in the superconducting (SC) state extrapolated to zero temperature increases with an increase of the hole concentration, $p$, in the overdoped regime.~\cite{Tallon, Loram}
These results indicate that the number of quasiparticles increases with increasing $p$ even in the SC ground state, suggesting the occurrence of a phase separation into SC and normal-state regions in the overdoped high-$T_{\rm c}$ cuprates. 
The phase separation in the overdoped regime has also been suggested by transverse-field muon-spin-relaxation measurements of TBCO~\cite{uemura,niedermayer} and Y$_{0.8}$Ca$_{0.2}$Ba$_2$Cu$_{3-z}$Zn$_z$O$_{7-\delta}$ ~\cite{bernhard} revealing that the muon-spin depolarization rate proportional to the SC carrier density decreases with increasing $p$  and by nuclear-magnetic-resonance measurements of LSCO revealing that the residual spin Knight shift in the SC ground state increases with increasing $p$.~\cite{Ohsugi}
Very recently, we have investigated the possible phase separation in the overdoped regime through the estimation of the SC volume fraction of LSCO from measurements of the magnetic susceptibility, $\chi$, on field cooling.~\cite{Tanabe-1,Tanabe-2,Adachi,Tanabe-3}
As a result, it has been found that the absolute value of $\chi$ at 2 K on field cooling decreases with an increase of $x$.
Therefore, it has been concluded that the SC volume fraction decreases with increasing $x$, supporting the occurrence of the phase separation into SC and normal-state regions in the overdoped regime of LSCO.

The next issue is whether the phase separation is as microscopic as suggested from the scanning-tunneling-microscopy measurements~\cite{pan} or as macroscopic as being comparable to the penetration depth of a few thousand angstrom.
In a microscopically phase-separated state, weak SC regions may appear around the boundary between intrinsic SC and normal-state regions due to the proximity effect.
In this case, the application of a moderate magnetic field may bring about the destruction of the weak superconductivity, producing pinning centers for vortices in the CuO$_2$ plane.
Therefore, an enhancement of the vortex pinning by the application of a moderate magnetic field may be detected by measuring $\chi$ or the magnetization, $M$.
On the other hand, in a macroscopically phase-separated state, weak SC regions between intrinsic SC and normal-state regions do not become dominant.
In this case, no marked enhancement of the vortex pinning due to the destruction of the weak SC regions is expected.
In this paper, we have measured the temperature dependence of $\chi$ in various magnetic fields on warming after zero-field cooling and also measured the magnetization curve, $M$ vs. $H$, up to 7 T at various temperatures in the overdoped regime of LSCO, in order to clarify whether the scale of the phase separation is microscopic or macroscopic.

\section{Experimental details}
Single crystals of LSCO were grown by the traveling-solvent floating-zone (TSFZ) method under flowing O$_{\rm 2}$ gas of 4 or 9 bar.
The details of the preparation of powders for the feed and solvent rods and the single-crystal growth have been reported elsewhere.~\cite{Tanabe-1,kawamata}
As-grown single-crystal rods were annealed in flowing O$_{\rm 2}$ gas of 1 bar at 900 $^{\rm o}$C for 50 h, cooled down to 500 $^{\rm o}$C at a rate of 8 $^{\rm o}$C/h, kept at 500 $^{\rm o}$C for 50 h and then cooled down to room temperature at a rate of 8 $^{\rm o}$C/h. 
The quality of the single crystals was checked by the x-ray back-Laue photography to be good. 
The crystals were also checked by the powder x-ray diffraction to be a single phase. 
The chemical composition of the single crystals was analyzed by the inductively coupled plasma optical emission spectrometry (ICP-OES). 
The distribution of the Sr content was also checked using an electron probe microanalyzer (EPMA) to be homogeneous in a crystal within the experimental accuracy.
The oxygen deficiency $\delta$ in La$_{2-x}$Sr$_x$CuO$_{4-\delta}$ was estimated from the iodometric titration to be 0.014 $\pm$ 0.01 and almost identical to each other for the crystals of $x$ = 0.178 - 0.261. 
For the $\chi$ vs. $T$ and $M$ vs. $H$ measurements, bulk single crystals were formed into the same rectangular shape of 1.82 mm and 0.68 mm in the ab-plane and 0.98 mm along the c-axis with the error of $\pm$ 5 \% to make the demagnetizing-field effect identical to each other. 
Both $\chi$ vs. $T$ and $M$ vs. $H$ measurements were carried out in magnetic fields up to 7 T at low temperatures down to 2 K, using a superconducting quantum interference device (SQUID) magnetometer (Quantum Design, MPMS-XL7).

\section{Results}\label{results}
\begin{figure}[tbp]
\begin{center}
\includegraphics[width=1.0\linewidth]{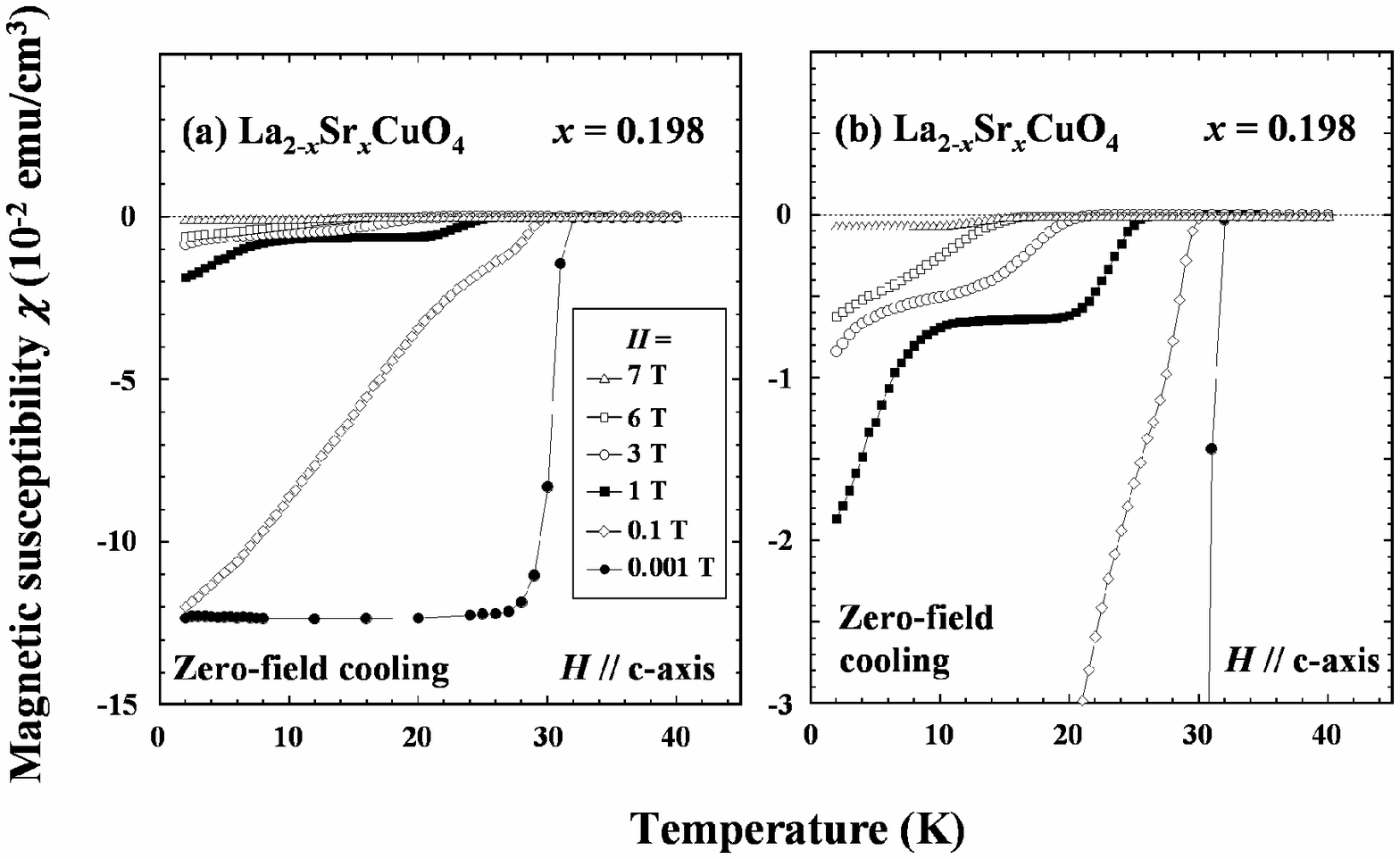}
\end{center}
\caption{(a) Temperature dependence of the magnetic susceptibility, $\chi$, for La$_{2-x}$Sr$_x$CuO$_4$ with $x = 0.198$ in magnetic fields of 0.001 T $\leq$ $H$ $\leq$ 7 T parallel to the c-axis on warming after zero-field cooling. (b) Magnified plots of $\chi$ in (a).}
\label{x=0.198ZFC} 
\end{figure}

Figure \ref{x=0.198ZFC} shows the temperature dependence of $\chi$ in magnetic fields of 0.001 T $\leq H \leq$ 7 T on warming after zero-field cooling in LSCO with $x$ = 0.198. 
The SC transition in a field of 0.001 T below the lower critical field, $H_{\rm c1}$, is sharp suggesting the good quality of the crystal.
With increasing field, the SC transition becomes broad, but a clear two-step transition is observed in 1 T.
In high magnetic fields above 1 T, the two-step transition tends to be smeared out with increasing field and changes to a single broad one in 7 T.
In the case of $\chi$ vs. $T$ on field cooling, on the other hand, the SC transition tends to become broad monotonically with increasing field and no two-step transition is observed.

\begin{figure}[tbp]
\begin{center}
\includegraphics[width=0.85\linewidth]{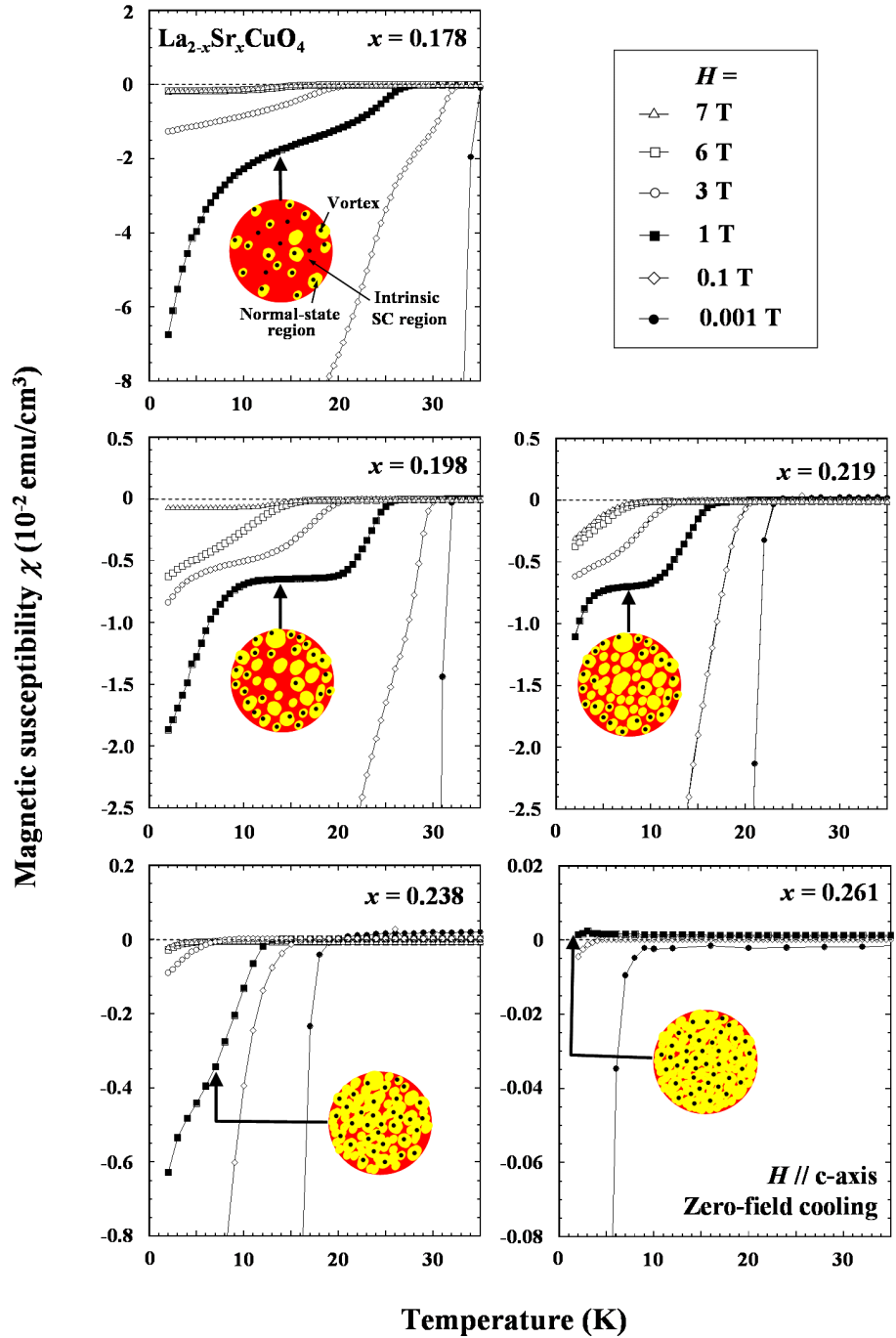}
\end{center}
\caption{(Color online) Temperature dependence of the magnetic susceptibility, $\chi$, for La$_{2-x}$Sr$_x$CuO$_4$ with $x = 0.178 - 0.261$ in magnetic fields of 0.001 T $\leq H \leq$ 7 T parallel to the c-axis on warming after zero-field cooling. Insets are schematic figures of the expected spatial distribution of vortices in the CuO$_2$ plane at moderate temperatures shown by arrows.}
\label{10kOeZFC} 
\end{figure}

Figure \ref{10kOeZFC} shows the temperature dependence of $\chi$ in magnetic fields of 0.001 T $\leq H \leq$ 7 T on warming after zero-field cooling for LSCO with $x$ = 0.178 - 0.261.
The SC transition shows a weak two-step feature for $x$ = 0.178 and shows a clear two-step transition for $x$ = 0.198 and 0.219.
On the other hand, the two-step transition tends to be smeared out for $x$ $\geq$ 0.238.
It is noted that no two-step SC transition has been observed in similar $\chi$ vs. $T$ measurements for a bulk single crystal of NbSe$_2$ which is regarded as a conventional homogeneous superconductor.~\cite{NbSe2}
Therefore, the two-step SC transition observed in LSCO is probably a characteristic feature of the overdoped high-$T_{\rm c}$ cuprates.

\begin{figure}[tbp]
\begin{center}
\includegraphics[width=1.0\linewidth]{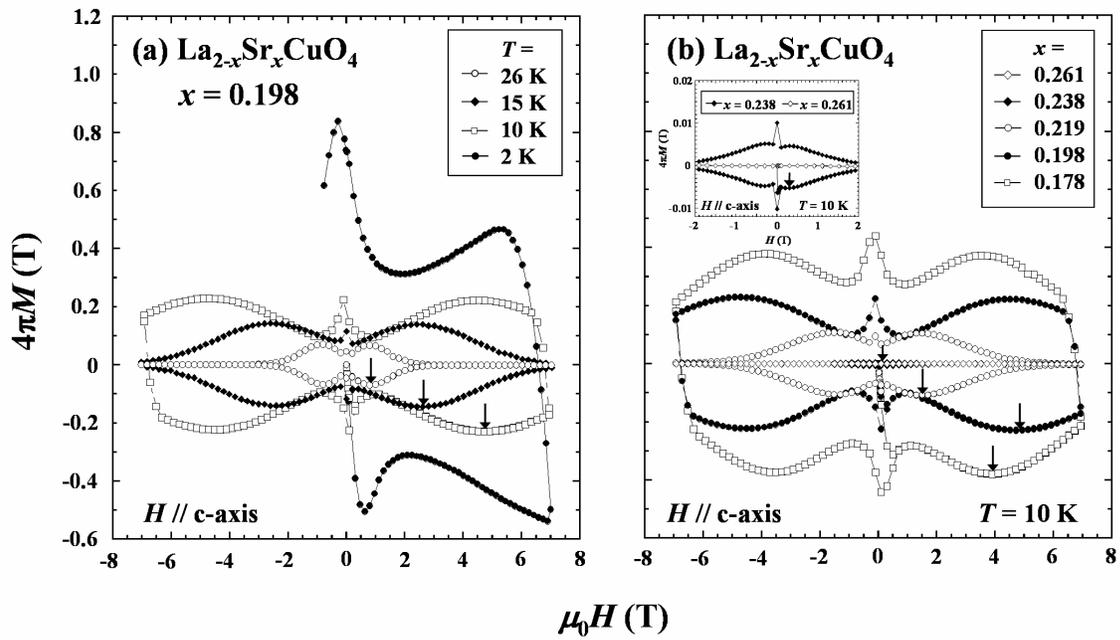}
\end{center}
\caption{Magnetization curves, $M$ vs. $H$, parallel to the c-axis up to 7 T for La$_{2-x}$Sr$_x$CuO$_4$ (a) with $x$ = 0.198 at temperatures of 2 K $\leq$ $T$ $\leq$ 26 K and (b) with 0.178 $\leq$ $x$ $\leq$ 0.261 at 10 K. The inset in (b) shows magnified plots of $M$ vs. $H$ for $x$ = 0.238 and 0.261. Arrows indicate so-called second peaks.}
\label{M-Hx=0.198} 
\end{figure}

Figures \ref{M-Hx=0.198}(a) and (b) show $M$ vs. $H$ for LSCO with $x$ = 0.198 at temperatures of 2 K $\leq T \leq$ 26 K and that with 0.178 $\leq x \leq$ 0.261 at 10 K, respectively.
For $x$ = 0.198, a so-called second peak due to a large hysteresis of $M$ vs. $H$ in magnetic fields between $H_{c1}$ and the upper critical field, $H_{c2}$, is observed at all measured temperatures.
The second peak indicates an increase of the vortex pinning, as typically observed in commercial bulk materials of REBa$_2$Cu$_3$O$_{7-\delta}$ including a small amount of the second phase of RE$_2$BaCuO$_5$.~\cite{REbulk1, REbulk2} 
In Fig. \ref{M-Hx=0.198}(a), it is found that $M$ vs. $H$ curves at 10 K and 15 K overlap each other in magnetic fields around 1 - 2 T, which is consistent with the result that $\chi$ vs. $T$ in 1 T is almost independent of temperature between 10 K and 20 K as shown in Figs. \ref{M-Hx=0.198}(a) and (b).
Moreover, $H$ = 1 T at 10 K in $M$ vs. $H$ is the onset field above which the vortex-pinning effect becomes marked, and $T$ = 10 K in 1 T in $\chi$ vs. $T$ is the onset temperature above which $\chi$ is almost independent of temperature.
In Fig. \ref{M-Hx=0.198}(b), it is found that the second peak is also observed for 0.178 $\leq$ $x$ $\leq$ 0.238 where the two-step SC transition is observed in greater or less degree in $\chi$ vs. $T$ as shown in Fig. \ref{10kOeZFC}.
These results indicate that the two-step SC transition in $\chi$ vs. $T$ is well correlated with the second peak in $M$ vs. $H$.
\section{Discussion}
The clear two-step SC transition in $\chi$ vs. $T$ on warming after zero-field cooling observed for LSCO with $x$ = 0.198 and 0.219 indicates that the penetration of vortices into the crystal is blocked in spite of warming in a moderate temperature-range.
Considering the correlation with the second peak in $M$ vs. $H$, these results are understood as follows. That is, strong vortex-pinning takes place near the surface of the crystal in a moderate temperature-range on warming for $x$ = 0.198 and 0.219 so that the magnetic field can not penetrate into the inside of the crystal, leading to the appearance of a plateau in $\chi$ vs. $T$.
It is noted that, although the second peak in $M$ vs. $H$ has been observed in some high-$T_{\rm c}$ cuprates,~\cite{nishizaki3}  such almost temperature-independent $\chi$ vs. $T$ in a wide temperature-range as in the present case has never been observed to date, to our knowledge, suggesting an unusual mechanism of the vortex pinning in the overdoped LSCO.~\cite{nishizaki,nishizaki2}

Formerly, it has been pointed out that a second peak appears in $M$ vs. $H$ in the overdoped LSCO and that it originates from the existence of oxygen defects.~\cite{O2diff-LSCO1,O2diff-LSCO2} 
That is, it has been supposed that oxygen defects bring about local destruction of the superconductivity, producing weak SC regions around themselves due to the proximity effect.
With increasing field, the weak SC regions change into normal-state ones earlier than intrinsic SC regions to operate as pinning centers for vortices, resulting in an enhancement of the vortex pinning.
In the present case, however, the oxygen deficiency $\delta$ is as small as 0.014 $\pm$ 0.01 and almost identical to one another for the crystals of $x$ = 0.178 - 0.261.
Moreover, the strong vortex-pinning is realized in a moderate temperature-range for $x$ = 0.198 and 0.219 and the vortex pinning becomes weak for $x$ $\geq$ 0.238.
Therefore, this scenario based upon only oxygen defects can not explain the present results.
Moreover, it is noted that the scenario based upon the bad crystallinity for $x$ = 0.198 and 0.219 can not explain the present results also, because neither marked increase in the width of the x-ray diffraction peaks nor marked increase in the normal-state in-plane electrical resistivity at low temperatures were observed around $x$ = 0.198 - 0.219.

\begin{figure}[tbp]
\begin{center}
\includegraphics[width=1.0\linewidth]{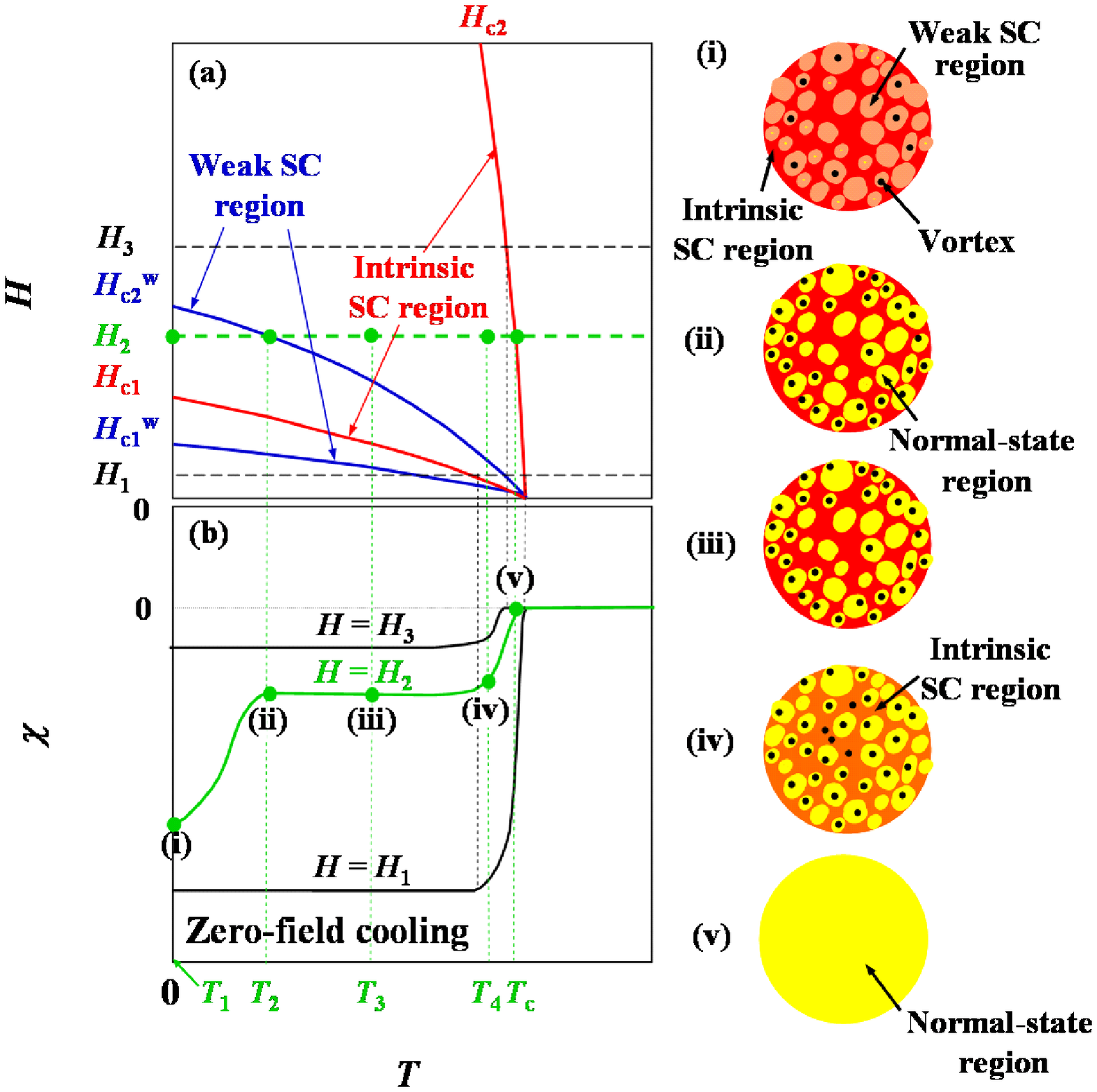}
\end{center}
\caption{(Color online) Schematic figures of (a) the $H$ - $T$ phase diagram and (b) the temperature dependence of $\chi$ on warming after zero-field cooling in a microscopically phase-separated state in $H$ = $H_1$, $H_2$, $H_3$. $H_{\rm c1}$ and $H_{\rm c2}$ are the lower and upper critical fields in intrinsic SC regions, respectively. $H_{\rm c1}^{\rm w}$ and $H_{\rm c2}^{\rm w}$ are the lower and upper critical fields in weak SC regions, respectively. Right figures (i) - (v) show the expected spatial distribution of vortices in the CuO$_2$ plane at (i) - (v) in (b), respectively.}
\label{H-T} 
\end{figure}

Here, we propose a possible scenario based upon a microscopic phase separation into SC and normal-state regions to explain the present results.
Figures \ref{H-T}(a) and (b) schematically show the $H$ - $T$ phase diagram and the temperature dependence of $\chi$ on warming after zero-field cooling in a microscopically phase-separated state, respectively.
In a microscopically phase-separated state, weak SC regions appear around the boundary between intrinsic SC and normal-state regions due to the proximity effect.
Supposed that microscopic weak SC regions are ubiquitously distributed in a crystal, the superconductivity in weak SC regions tends to be destroyed earlier than that in intrinsic SC regions with increasing temperature or field, so that the weak SC regions change to normal-state regions regarded as pinning centers for vortices.
In the case of Fig. \ref{H-T}(i), the applied magnetic field, $H_2$, is between $H_{\rm c1}$ in the intrinsic SC regions and the upper critical field in the weak SC regions, $H_{\rm c2}^{\rm w}$, at the lowest temperature, $T_1$.
In this case, the number of vortices penetrating into the crystal increases with increasing temperature, resulting in the decrease of the shielding effect of superconducting currents. 
When the temperature reaches $T_2$ in Fig. \ref{H-T}, a number of microscopic normal-state regions appear in the crystal due to the destruction of the superconductivity in the weak SC regions, as shown in Fig. \ref{H-T}(ii).
In this case, the magnetic flux coming from the outside of the crystal is pinned in the normal-state regions near the surface of the crystal.
With further increasing temperature, the magnetic flux coming from the outside of the crystal is still pinned in the normal-state regions near the surface of the crystal, as shown in Fig. \ref{H-T}(iii), due to the large SC condensation energy of the intrinsic SC regions, leading to the appearance of a plateau in $\chi$ vs. $T$.  
When the temperature reaches $T_4$ in Fig. \ref{H-T}, vortices enter the inside of the crystal, because the SC condensation energy in the intrinsic SC regions decreases at temperatures near the SC transition temperature, $T_{\rm c}$. This results in the decrease and disappearance of the shielding effect, as shown in Figs. \ref{H-T}(iv) and (v).
This kind of behavior of $\chi$ vs. $T$ is clearly observed for $x$ = 0.198 and 0.219. 

According to this scenario, the $x$ dependence of the distribution of vortices in the CuO$_2$ plane is also understood as shown in insets of Fig. \ref{10kOeZFC}.
There, the penetration of vortices into the inside of the crystal is schematically shown for $x$ = 0.178 - 0.261, supposing that normal-state regions appear due to the destruction of weak SC regions on warming in a microscopically phase-separated state.
For $x$ = 0.178, the SC volume fraction is much larger than those of $x$ = 0.198 and 0.219.
In this case, vortices can penetrate into the inside of the crystal due to a small number of pinning centers for vortices, resulting in the weak two-step feature in $\chi$ vs. $T$.
Once normal-state regions become dominant with increasing $x$ for $x$ $\geq$ 0.238, on the other hand, neither normal-state nor weak SC regions act as pinning centers for vortices, which is consistent with the disappearance of the two-step feature in $\chi$ vs $T$ for $x$ $\geq$ 0.238 in which the SC volume fraction is much smaller than that of $x$ = 0.219.

It has already been reported from the magnetization measurements that the vortex-pinning effect is weak on account of the small SC volume fraction in the overdoped regime of LSCO, TBCO, Bi$_2$Sr$_2$CuO$_{6+\delta}$, Y$_{1-x}$Ca$_x$Ba$_2$Cu$_3$O$_{7-\delta}$, Bi$_2$Sr$_2$CaCu$_2$O$_{8+\delta}$.~\cite{wen-euro}	 
Therefore, the new important information from the present results is that a $microscopic$ phase separation into SC and normal-state regions takes place in the overdoped regime of LSCO, because the present results can hardly be explained if the phase separation is macroscopic.
Finally, it is noted that the strong pinning effect is not directly related to the crystal structure, because the structures of $x$ = 0.198 and 0.219 in the SC state are different from each other; the former is orthorhombic, while the latter is tetragonal.

\section{Summary}
In summary, it has been found from $\chi$ vs. $T$ measurements in magnetic fields parallel to the c-axis up to 7 T on warming after zero-field cooling in the overdoped regime of LSCO single crystals that $\chi$ is independent of temperature in a moderate temperature-range in the SC state in 1 T for $x$ = 0.198 and 0.219, while the almost temperature-independent $\chi$ disappears for $x$ $\geq$ 0.238.
Moreover, a second peak has markedly appeared in $M$ vs. $H$ measurements in the overdoped regime of LSCO.
These results indicate an anomalous enhancement of the vortex pinning and are understood assuming the occurrence of a $microscopic$ phase separation into SC and normal-state regions in the overdoped regime. 
That is, microscopic weak SC regions appear around the boundary between intrinsic SC regions and normal-state regions due to the proximity effect, and the superconductivity of the weak SC regions is destroyed earlier than that of the intrinsic SC regions with increasing temperature or field so that the weak SC regions operate as strong pinning centers for vortices, resulting in strong vortex pinning in a moderate range of temperature or field in a microscopically phase-separated SC state.
 Accordingly, these results strongly suggest that a $microscopic$ phase separation into SC and normal-state regions takes place in the overdoped high-$T_{\rm c}$ cuprates.

\section*{Acknowledgments}
We are indebted to K. Takada and M. Ishikuro for their help in the ICP-OES analysis. 
The EPMA analysis was supported by Y. Murakami in the Advanced Research Center of Metallic Glasses, Institute for Materials Research (IMR), Tohoku University and K. Kudo in IMR, Tohoku University.
Fruitful discussions with T. Nishizaki and S. Awaji are gratefully acknowledged. 
The $\chi$ measurements were carried out at the Center for Low Temperature Science, Tohoku University. This work was supported by the Iketani Science and Technology Foundation and also by a Grant-in-Aid for Scientific Research from the Ministry of Education, Culture, Sports, Science and Technology, Japan.

\begin{thebibliography}{99}
\bibitem{nishizaki}
A two-step SC transition in $\chi$ vs. $T$ has been reported for YBa$_2$Cu$_3$O$_{6.9}$ at high temperatures just below the irreversibility temperature, but still $\chi$ weakly depends on temperature.
This has been explained as being due to vortex pinning near the first-order vortex-melting transition.~\cite{nishizaki2}
\end{thebibliography}
\end{document}